\documentclass[5p,twocolumn]{elsarticle}
\usepackage{graphicx}	
\usepackage{amsmath}	
\usepackage{amssymb}	
\usepackage{physics}
\usepackage{listings}
\usepackage{xcolor}
\usepackage{lineno}
\usepackage{hyperref}
\usepackage{cleveref}[2012/02/15]

\journal{Astronomy and Computing}
\modulolinenumbers[5]
\biboptions{authoryear}

\lstdefinelanguage{toml}
{
  sensitive=false,
  morestring=[b]"
}

\definecolor{mystring}{RGB}{245, 121, 58}
\definecolor{mykey}{RGB}{169, 90, 161}
\definecolor{myheader}{RGB}{133, 192, 249}

\crefformat{footnote}{#2\footnotemark[#1]#3}

\defcitealias{hera_opm}{ascl:2104.001}
\defcitealias{librarian}{ascl:2104.002}

\begin{document}

\begin{frontmatter}

\title{A Real Time Processing System for Big Data in Astronomy: Applications to HERA}

\author[ucb,bccp,penn]{Paul La Plante\corref{correspondingauthor}}
\cortext[correspondingauthor]{Corresponding Author}
\ead{plaplant@berkeley.edu}

\author[cfa,aas]{Peter K. G. Williams}
\author[asu]{Matthew Kolopanis}
\author[ucb]{Joshua S. Dillon\fnref{aapf}}
\author[asu,winona]{Adam P. Beardsley\fnref{aapf}}
\author[ucb,mit]{Nicholas S. Kern}
\author[uw]{Michael Wilensky}
\author[ucb]{Zaki S. Ali}
\author[ucb]{Zara Abdurashidova}
\author[penn]{James E. Aguirre}
\author[cambridge]{Paul Alexander}
\author[sarao]{Yanga Balfour}
\author[rhodes,inaf,sarao]{Gianni Bernardi}
\author[penn]{Tashalee S. Billings}
\author[asu]{Judd D. Bowman}
\author[nraoc]{Richard F. Bradley}
\author[qmul]{Phil Bull}
\author[brown]{Jacob Burba}
\author[cambridge]{Steve Carey}
\author[nraos]{Chris L. Carilli}
\author[ucb]{Carina Cheng}
\author[ucb]{David R. DeBoer}
\author[ucb]{Matt Dexter}
\author[cambridge]{Eloy de Lera Acedo}
\author[cambridge]{John Ely}
\author[ucb]{Aaron Ewall-Wice}
\author[cambridge]{Nicolas Fagnoni}
\author[sarao]{Randall Fritz}
\author[ucla]{Steven R. Furlanetto}
\author[cambridge]{Kingsley Gale-Sides}
\author[nraos]{Brian Glendenning}
\author[ucb]{Deepthi Gorthi}
\author[melbourne]{Bradley Greig}
\author[sarao]{Jasper Grobbelaar}
\author[sarao]{Ziyaad Halday}
\author[uw,escience]{Bryna J. Hazelton}
\author[mit]{Jacqueline N. Hewitt}
\author[ucb]{Jack Hickish}
\author[asu]{Daniel C. Jacobs}
\author[sarao]{Austin Julius}
\author[brown]{Joshua Kerrigan}
\author[uwc]{Piyanat Kittiwisit}
\author[penn]{Saul A. Kohn}
\author[brown]{Adam Lanman}
\author[sarao]{Telalo Lekalake}
\author[asu]{David Lewis}
\author[mcgill]{Adrian Liu}
\author[ucb]{David MacMahon}
\author[sarao]{Lourence Malan}
\author[sarao]{Cresshim Malgas}
\author[sarao]{Matthys Maree}
\author[penn]{Zachary E. Martinot}
\author[sarao]{Eunice Matsetela}
\author[sns]{Andrei Mesinger}
\author[sarao]{Mathakane Molewa}
\author[uw]{Miguel F. Morales}
\author[sarao]{Tshegofalang Mosiane}
\author[asu]{Steven Murray}
\author[mit]{Abraham R. Neben}
\author[cambridge]{Bojan Nikolic}
\author[ucb]{Aaron R. Parsons}
\author[ucb,mcgill]{Robert Pascua}
\author[ucb]{Nipanjana Patra}
\author[sarao]{Samantha Pieterse}
\author[brown]{Jonathan C. Pober}
\author[cambridge]{Nima Razavi-Ghods}
\author[uw]{Jon Ringuette}
\author[nraos]{James Robnett}
\author[sarao]{Kathryn Rosie}
\author[sarao,uwc]{Mario G. Santos}
\author[brown]{Peter Sims}
\author[sarao]{Craig Smith}
\author[sarao]{Angelo Syce}
\author[nraos]{Nithyanandan Thyagarajan\fnref{jansky}}
\author[mit]{Haoxuan Zheng}

\address[ucb]{Department of Astronomy, University of California, Berkeley, CA}
\address[bccp]{Berkeley Center for Cosmological Physics, University of California, Berkeley, CA}
\address[penn]{Department of Physics and Astronomy, University of Pennsylvania, Philadelphia, PA}
\address[cfa]{Center for Astrophysics \textbar{} Harvard \& Smithsonian, Cambridge, MA}
\address[aas]{American Astronomical Society, Washington, DC}
\address[asu]{School of Earth and Space Exploration, Arizona State University, Tempe, AZ}
\address[winona]{Physics Department, Winona State University, Winona, MN}
\address[mit]{Department of Physics, Massachusetts Institute of Technology, Cambridge, MA}
\address[uw]{Department of Physics, University of Washington, Seattle, WA}
\address[cambridge]{Cavendish Astrophysics, University of Cambridge, Cambridge, UK}
\address[sarao]{South African Radio Astronomy Observatory, Cape Town, South Africa}
\address[rhodes]{Department of Physics and Electronics, Rhodes University, Grahamstown, South Africa}
\address[inaf]{INAF-Instituto di Radioastronomia, Bologna, Italy}
\address[nraoc]{National Radio Astronomy Observatory, Charlottesville, VA}
\address[qmul]{Queen Mary University London, London, UK}
\address[brown]{Department of Physics, Brown University, Providence, RI}
\address[nraos]{National Radio Astronomy Observatory, Socorro, NM}
\address[ucla]{Department of Physics and Astronomy, University of California, Los Angeles, CA}
\address[melbourne]{Department of Physics, University of Melbourne, Parkville, VIC 2010, Australia}
\address[escience]{eScience Institute, University of Washington, Seattle, WA}
\address[uwc]{Department of Physics and Astronomy, University of Western Cape, Cape Town, South Africa}
\address[mcgill]{Department of Physics and McGill Space Science Institue, McGill University, Montreal, Canada}
\address[sns]{Scuola Normale Superiore, 56126 Pisa, PI, Italy}

\fntext[aapf]{NSF Astronomy and Astrophysics Postdoctoral Fellow}
\fntext[jansky]{NRAO Jansky Fellow}

\begin{abstract}
  As current- and next-generation astronomical instruments come online, they
  will generate an unprecedented deluge of data. Analyzing these data in real
  time presents unique conceptual and computational challenges, and their
  long-term storage and archiving is scientifically essential for generating
  reliable, reproducible results. We present here the real-time processing (RTP)
  system for the Hydrogen Epoch of Reionization Array (HERA), a radio
  interferometer endeavoring to provide the first detection of the highly
  redshifted 21\,cm signal from Cosmic Dawn and the Epoch of Reionization by an
  interferometer. The RTP system consists of analysis routines run on raw data
  shortly after they are acquired, such as calibration and detection of
  radio-frequency interference (RFI) events. RTP works closely with the
  Librarian, the HERA data storage and transfer manager which automatically
  ingests data and transfers copies to other clusters for post-processing
  analysis. Both the RTP system and the Librarian are public and open source
  software, which allows for them to be modified for use in other scientific
  collaborations. When fully constructed, HERA is projected to generate over 50
  terabytes (TB) of data each night, and the RTP system enables the successful
  scientific analysis of these data.
\end{abstract}


\end{frontmatter}




\section{Introduction}

\label{sec:intro}
In recent years, the amount of scientific data has exploded. Astronomy is no
exception to this transformation, and experiments are generating more data than
at any point in the past. The acceleration of data generation has outpaced
Moore's Law for storage, causing data \textit{transfer} to join data
\textit{storage} among the technical challenges that researchers must grapple
with. These closely related computational requirements of analyzing data and
storing them are near-universal problems that must be overcome to enable the
scientific goals of these experiments. Indeed, as the size and number of files
continues to grow, the very process of handling the data becomes non-trivial to
solve and can rival the sophistication of the actual analysis being performed.

The field of radio astronomy boasts some of the highest data rates and volumes
in all of astronomy and astrophysics research. Digitization of the radio
spectrum (as a function of time, frequency, and instrumental polarization
component) is close to an ``embarrassingly parallel'' problem, so that the data
rate out of a radio telescope essentially is only bounded by its budget for
digital signal processing (DSP) hardware. Furthermore, many radio telescopes are
interferometers, which cross-correlate the signals measured by pairs of
receivers, yielding a total data rate that scales as the square of the number of
receivers. The computer clusters that perform interferometric cross-correlations
are among the largest single-purpose machines built for scientific research and
have some of the highest sustained-throughput data rates in the world. Radio
observatories aim to operate with high duty cycles, generally 12--24 hours a
day, such that their large data rates lead to large data volumes as well.

Once data have been taken, compute-intensive tasks such as calibration and the
excision of radio frequency interference (RFI) generally must be performed
before scientific analysis can proceed. Due to the high duty cycles of typical
radio observatories, these algorithms must run reliably in close to real time, a
substantial challenge at the data rates of current-generation experiments such
as the Hydrogen Epoch of Reionization Array
(HERA\footnote{\url{https://reionization.org}}, \cite{deboer_etal2017}), let
alone next-generation ones such as the Square Kilometre Array
(SKA\footnote{\url{https://skatelescope.org}}). The need for real-time analysis
is further enhanced because the data rates of these experiments are such that it
is computationally or financially unfeasible to store the full raw
data. Immediate data reduction is necessary to attain data volumes that can be
transferred or archived.

In order to deal with the ever-increasing demands of data storage and
processing, the HERA collaboration has developed the Real Time Processing (RTP)
system and the Librarian, which together support the analysis and storage of raw
and reduced HERA data products. The RTP system is built on a Python package
called
\texttt{hera\_opm}\footnote{\label{fn:rtp}\url{https://github.com/HERA-Team/hera_opm}}
\citepalias{hera_opm}---the HERA Online Processing Module---which defines and
manages a workflow of analysis steps. At the same time, RTP is also composed of
various monitoring systems that track the progress of analysis steps, with an
eye toward automatic diagnosis of potential processing issues. The
Librarian\footnote{\label{fn:librarian}\url{https://github.com/HERA-Team/librarian}}
\citepalias{librarian} is also a Python package and supports long-term storage
of raw and processed data files and facilitates movement of data between
different computing facilities. Both of these packages are publicly available
and licensed under open-source licenses, with the hope that they may be widely
useful to the broader astronomical research community and beyond.

Although in this paper we discuss these systems primarily in the context of
HERA, the frameworks are sufficiently general that they may be adapted to other
purposes with relatively little modification. For example, the Librarian has
been adopted by the Simons Observatory
(SO\footnote{\url{https://simonsobservatory.org}}), and the primary
functionality of \texttt{hera\_opm} does not reference specifics of HERA data
files or analysis techniques. The RTP and the Librarian systems have been built
to be modular, which allows for them to be adopted to use in other contexts. At
the same time, their well-documented and reliably tested codebase can provide a
stable platform on which to build, without the need to reinvent existing
infrastructure. The paper below is organized as follows: in
Sec.~\ref{sec:requirements}, we outline the data processing and storage
requirements relevant for HERA. In Sec.~\ref{sec:rtp}, we describe the RTP
system and present examples of its use. In Sec.~\ref{sec:librarian}, we describe
the Librarian system. In Sec.~\ref{sec:summary} we provide additional
discussion and describe future directions of these systems.

\section{Processing and Storage Requirements}
\label{sec:requirements}
In principle, an interferometer can measure a full correlation matrix of
cross-correlated antenna signals for a series of times and frequencies. This is
a significant amount of data which must be moved, calibrated, imaged,
etc. Although in practice there are often cuts made to the number of antennas,
times, and/or frequencies used in data analysis schemes, maximizing the
sensitivity and scientific return of an experiment necessitates handling as much
of the data as possible in a reliable and efficient manner.

An array with $N_\mathrm{ant}$ elements has $N_\mathrm{ant}(N_\mathrm{ant}+1)/2$
unique pairs of antennas (including auto-correlations). Each of these baselines
produces a spectrum consisting of $N_\mathrm{freq}$ frequency channels, which is
produced by the correlator every $\Delta t_\mathrm{corr}$ seconds. Such a
spectrum is produced for each polarization element of the antenna, and
cross-correlated when forming visiblities to produce $N_\mathrm{pol}$
instrumental polarization spectra. For a given observation window
$T_\mathrm{obs}$, the total number of spectra generated is
$N_\mathrm{time} = T_\mathrm{obs}/\Delta t_\mathrm{corr}$. Visibilities are
typically recorded as single-precision floating-point complex numbers, where the
real and imaginary components each require 4 bytes to store. Thus, the total
volume of data $V$ produced by HERA in a single night of observation is:
\begin{equation}
V = \frac{1}{2} N_\mathrm{ant}(N_\mathrm{ant}+1)N_\mathrm{freq}N_\mathrm{pol}N_\mathrm{time} \times 2 \times 4\,\mathrm{bytes}.
\label{eqn:data_vol}
\end{equation}
The fully constructed HERA array will consist of $N_\mathrm{ant} = 350$,
$N_\mathrm{freq} = 6144$, $N_\mathrm{pol} = 4$, and produce spectra every 2
seconds. For a typical observation of 12 hours ($N_\mathrm{time} = 21,600$),
this leads to a raw data volume of over 260 terabytes (TB). The actual data
volume recorded by HERA is significantly smaller than this (e.g., using baseline
dependent averaging, as described in \citealt{wijnholds_etal2018}), though the
resulting data volume per night is projected to be well over 50 TB. A typical
observing season for HERA is 100 days, so several petabytes of data are
generated within a single observing season.

Like many other radio observatories, HERA is situated in a remote location to
avoid terrestrial sources of RFI. In this case, HERA is in the Karoo desert of
South Africa, hosted by the South African Radio Astronomy Observatory
(SARAO\footnote{\url{https://www.sarao.ac.za/}}). This means that both on-site
computational resources and network bandwidth to the wider world are limited,
raising further challenges. A representative bandwidth of 100 Mbps is only
capable of moving about 1 TB per day, and so significant processing of raw data
must be done to facilitate adequately small data products. Because HERA is
designed to observe at a $\sim$50\% duty cycle (12 hours every night), the RTP
system must be able to do its work within 24 hours to avoid falling behind on
data analysis. This requirement adds time pressure to the RTP system's
processing requirement, as the raw data cannot be stored indefinitely.
Accordingly the RTP system monitors files being processed and identifies
potential blockages of the pipeline. Finally, data products of sufficient
quality must be produced to allow HERA to achieve its science goals, which
requires a dynamic range between foreground signals and the expected
cosmological signal of $\sim10^5$ \citep{pober_etal2013}.

\subsection{Design Considerations}
\label{sec:system_design}
Before discussing the RTP and the Librarian systems in detail, we begin by
outlining some of the basic requirements of the systems, and some of the
considerations driving the design of these systems. By laying out these ideas,
we motivate the choices made in the systems, and demonstrate the applicability
to other systems with similar applications.

For the RTP system, the main features are:
\begin{itemize}
\item enabling a way to easily define a workflow (i.e., fixed set of tasks) and
  apply it to an arbitrary set of input data;
\item allowing for flexibly interfacing with various cluster resource schedulers
  (e.g., Slurm, \footnote{\url{https://slurm.schedmd.com/}}
  TORQUE,\footnote{\url{https://adaptivecomputing.com/cherry-services/torque-resource-manager/}}
  etc.);
\item providing the option for the user to group together and operate on
  ``sets'' of files if desired (e.g., flagging of radio-frequency
    interference (RFI) may work better on longer ``chunks'' of data);
\item ensuring high uptime and communication with users about the status of the
  pipeline execution.
\end{itemize}

For the Librarian system, the main features are:
\begin{itemize}
\item providing a reliable way to save telescope data for long-term storage;
\item communicating between multiple Librarian servers and transferring data
  easily;
\item ensuring data integrity when replicating data across servers;
\item allowing for adding and managing multiple storage locations of a Librarian
  server;
\item supporting access by automated pipelines;
\item automating data movement within and between sites.
\end{itemize}

For both systems, we also want to be running based on a highly tested and
well-documented code base to provide confidence in scientific results. This
approach helps build confidence that the results are reliable.

Although the RTP and the Librarian have been developed expressly in the context
of HERA, there are other data management and transport systems that are used by
other observatories and experiments. One such system is the Next Generation
Archive System (NGAS\footnote{\url{https://github.com/ICRAR/ngas}}), which is
used for other radio astronomy experiments such as the Atacama Large Millimeter
Array (ALMA\footnote{\url{https://www.almaobservatory.org/en/home/}}). The
Murchison Widefield Array (MWA\footnote{\url{https://www.mwatelescope.org}})
uses a system known as
Manta-ray\footnote{\url{https://github.com/MWATelescope/manta-ray-client}},
which serves as a client for the data archiving and storage system. The Low
Frequency Array (LOFAR\footnote{\url{https://www.astron.nl/telescopes/lofar}})
hosts a ``Long Term Archive'' of 43 petabytes (PB) which supports archiving and
retrieval of data for scientific analysis \citep{spreeuw_etal2019}. Outside of
the field of radio astronomy, large collaborations like the Large Hadron
Collider have developed systems for handling enormous data volumes, and devised
plans for dealing with these issues in future research endeavors
\citep{cern2017}. However, none of these solutions quite fit the use needs of
the HERA collaboration: the source code was sometimes proprietary, and
installation was often difficult. Other solutions are primarily for a long-term
archive system that does not necessarily support frequent access for real-time
analysis. Thus, the need arose to build a series of tightly coupled
data-processing and data-storage systems to facilitate the needs of the
collaboration.

In addition, HERA is an official pathfinder of the SKA, whose data processing
and handling needs will be even more demanding than those of HERA. Although
there is no formal agreement between the collaborations, many of the
developments made throughout the course of the experiment can be used to inform
design decisions for the SKA. Additionally, having the RTP and Librarian code
bases as public repositories with documentation can provide important insights
to be used as more of the SKA system is built and developed.

\subsection{Terminology}
Throughout the rest of the discussion, we make use of several key words that
have special meaning to the RTP and the Librarian systems. We briefly summarize
these terms here, so that the reader may refer back to them later.
\begin{itemize}
\item \textit{Observation}: a single file generated by the HERA correlator
  system. Due to internal throughput constraints of the correlator, these tend
  to be 16 seconds in length, and are projected to be $\sim$25 GB in size when
  observing for 350 antennas.
\item \textit{Observing session}: the combined data product for a continuous set
  of \textit{observations}. Generally for HERA these comprise 10-12 hours of
  sustained observation. They are composed of roughly 2000 individual
  observations.
\item \textit{File}: in the context of the Librarian, this is an abstract
  definition of any data product. While it may correspond to a single file in
  the traditional ``filesystem-like'' sense, it may also be a directory treated
  as a single object.  For data generated by telescopes, they can be indexed by
  membership in an \textit{observation} or an \textit{observing session}. As
  discussed more below in Sec.~\ref{sec:lib_arch}, the name, size, and hash are
  required/guaranteed to be unique for a particular data product.
\item \textit{File instance}: in the context of the Librarian, this is a
  specific \textit{file} that is stored on a particular Librarian server. As
  discussed more below in Sec.~\ref{sec:lib_arch}, a Librarian server is allowed
  to have multiple copies of a given file instance, or may not have any local
  instances at all.
\end{itemize}
In the following discussion, we have emphasized these words when used in a
context that connotes the above meanings.

\section{Real Time Processing}
\label{sec:rtp}
The real time pipeline developed for the reduction of HERA data has emerged
after several iterations stretching back over years of development beginning
during the PAPER project.\footnote{One such example is the
  \href{https://github.com/HERA-Team/RTP}{RTP repository}, though the original
  codebase is no longer under active development.} In these various iterations
we have investigated ways to support a range of operations with notable data
processing needs. What emerged is a global system for managing large data
volumes (Librarian), a processing system for running workflows on data (RTP),
and a monitor and control system. Here we focus primarily on RTP but provide a
brief description of the other supporting elements.

Though the discussion here is specific to HERA, many of the tools developed and
employed are publicly available and applicable to current and future
observatories with intense data processing needs. By presenting the general
functionality of these systems, we hope that other collaborations may find these
tools useful for constructing real-time systems for data-intensive tasks.

The focus of this section is primarily on the software infrastructure developed
for running HERA data analysis. We do not focus as much on the actual analysis
steps being run, which include RFI excision and calibration of visibilities. For
a more in-depth discussion of these analysis steps, see
\citet{dillon_etal2020,kern_etal2020a,kern_etal2020b,h1c_limits,validation_paper}. Though
there is some discussion of the specific processing steps, the architecture
presented is sufficiently flexible to be adapted to other applications in a
relatively straightforward fashion.

There are several existing pipeline management packages that already exist (such
as SciLuigi,\footnote{\url{https://github.com/pharmbio/sciluigi}}
Jug,\footnote{\url{https://github.com/luispedro/jug}} and
COSMOS\footnote{\url{https://mizzou-cbmi.github.io/}}) so it is worth discussing
what features the RTP system provides that existing infrastructure does not.
The primary use-case this package addresses that others do not is the ability to
apply a series of steps to a list of input data. Specifically, this applies to
data files that are time-ordered, with ``neighbors'' that come either before or
after them. These individual files can have arbitrary time boundaries, which can
change from night-to-night depending on the start- and stop-times of the
observation. Although efforts are made to be aligned to a common grid in local
sidereal time (LST) to facilitate averaging data together from different nights,
the precise start and stop times otherwise can be arbitrary. For instance, these
times may change based on the time of year to avoid observing the sun, which can
lead to systematic errors in data analysis and temperature-related fluctuations
in the behavior of electronics.  Many algorithms, such as a time domain
convolution, are applied on a time range larger than the native length of a data
file.  RTP handles the concept of pre-requisite jobs in a user-friendly
manner. These pre-requisites specify tasks that must complete before a given
task is launched, usually because its output is required for a future
step. These pre-requisites can include a prior step in the workflow for the file
itself, or for an arbitrary number of time-neighbors either before or after
it. In addition to allowing the user to specify pre-requisites in time, the user
is also able to provide a partitioning of input files along the time axis to
pass to a specific task. This allows the user flexibility in handling file
processing for different steps with different requirements, where some steps may
require a particular number of input files from multiple times, but other steps
should operate on each file individually.

We discuss these features in more detail below. In
Sec.~\ref{sec:rtp_architecture}, we discuss the general RTP architecture and
designed use-case. In Sec.~\ref{sec:onsite}, we talk about its use in the
real-time system employed by HERA. In Sec.~\ref{sec:offsite}, we discuss
additional workflows implemented using the \texttt{hera\_opm} framework in
service of HERA data processing, and outline some of the flexibility offered by
the package. In Sec.~\ref{sec:monitoring}, we briefly discuss some of the
monitoring tools that the RTP system uses to keep track of the status of
processing.

\subsection{RTP Architecture}
\label{sec:rtp_architecture}

The RTP system oversees the successful execution of the HERA analysis
pipeline. RTP consists of a loosely coupled set of independent processes, as
well as monitoring daemons that track the progress of analysis in the system as
a whole. These monitoring systems interface with other systems on site (such as
the Monitoring \& Control
system\footnote{\url{https://github.com/HERA-Team/hera_mc}}). Many of the tools
that it relies on feature open-source licenses and are freely available to
download. Most of the core functionality is written in Python, which makes the
system highly portable. Additionally, many of these tools are
industry-standard. Typical clusters may already have them available to users or
they can be installed with little difficulty.

Once raw data are recorded by the correlator, the RTP system begins
operating. The raw data storage nodes are attached to the processing nodes via
network file system (NFS) mounted directories, which allows for accessing data
without explicitly copying files to local scratch space. Output files are also
written to this shared disk space. The analysis pipeline launches automatically
at the conclusion of observing each night without the need of human
intervention. Automation helps ensure that all data can be examined within the
time constraint, and so the RTP system automatically retries jobs in an attempt
to provide some robustness to occasional failures.

The primary goal of RTP is to run a series of compute-intensive analysis
tasks. For the workflow execution, it uses the \texttt{hera\_opm} package. This
package converts a user-defined workflow file (specified in the TOML
format\footnote{\url{https://github.com/toml-lang/toml}}, an example of which is
shown in Listing~\ref{list:config}) and a series of input files into concrete
steps in an overall work diagram. This diagram is written in a file format that
can be parsed by the \texttt{makeflow} package, which is available as part of
the
\texttt{cctools}\footnote{\url{https://github.com/cooperative-computing-lab/cctools}}
system \citep{makeflow1,makeflow2}. Internally, \texttt{makeflow} generates a
directed acyclic graph given dependencies for each step in the process, similar
to the \texttt{make} command line utility. Once this graph has been generated,
it manages the execution of each step in the process either locally or using
cluster scheduling systems. When using cluster management systems,
\texttt{makeflow} supports specifying batch job options, such as the number of
processing nodes and amount of memory required. There are also tools available
for verifying \texttt{makeflow} recipe files, as well as simple progress
monitoring tools. Taken together, this package represents a powerful method for
overseeing the execution of a given workflow, and serves as the backing pipeline
management tool for the \texttt{hera\_opm} package.

\begin{lstlisting}[float,frame=single,caption={A sample configuration file for the \texttt{hera\_opm} package.},label={list:config}]
[<@\color{myheader}Options@>]
<@\color{mykey}makeflow\_type@> = "analysis"
<@\color{mykey}path\_to\_do\_scripts@> = "/path/to/task_scripts"
<@\color{mykey}source\_script@> = "~/.bashrc_hera"
<@\color{mykey}conda\_env@> = "hera"
<@\color{mykey}base\_mem@> = 8000
<@\color{mykey}base\_cpu@> = 1
<@\color{mykey}timeout@> = "24h"

[<@\color{myheader}ANT\_METRICS\_OPTS@>]
<@\color{mykey}cross\_cut@> = 5.0
<@\color{mykey}dead\_cut@> = 5.0
<@\color{mykey}extension@> = ".ant_metrics.hdf5"

[<@\color{myheader}XRFI\_OPTS@>]
<@\color{mykey}kt\_size@> = 8
<@\color{mykey}kf\_size@> = 8

[<@\color{myheader}WorkFlow@>]
<@\color{mykey}actions@> = [
  "SETUP",
  "ANT_METRICS",
  "ADD_LIBRARIAN_ANT_METRICS",
  "XRFI",
  "ADD_LIBRARIAN_XRFI",
  "TEARDOWN"
]

[<@\color{myheader}ANT\_METRICS@>]
<@\color{mykey}args@> = [
  "{basename}",
  "${ANT_METRICS_OPTS:cross_cut}",
  "${ANT_METRICS_OPTS:dead_cut}"
]

[<@\color{myheader}ADD\_LIBRARIAN\_ANT\_METRICS@>]
<@\color{mykey}args@> = [
  "{basename}",
  "${ANT_METRICS_OPTS:extension}"
]
<@\color{mykey}prereqs@> = "ANT_METRICS"

[<@\color{myheader}XRFI@>]
<@\color{mykey}queue@> = "gpu"
<@\color{mykey}chunk\_size@> = 10
<@\color{mykey}stride\_length@> = 10
<@\color{mykey}time\_centered@> = true
<@\color{mykey}collect\_stragglers@> = true
<@\color{mykey}args@> = [
  "{basename}",
  "${XRFI_OPTS:kt_size}",
  "${XRFI_OPTS:kf_size}"
]

[<@\color{myheader}ADD\_LIBRARIAN\_XRFI@>]
<@\color{mykey}args@> = ["{basename}"]
<@\color{mykey}prereqs@> = "XRFI"
\end{lstlisting}

Listing~\ref{list:config} shows a sample TOML file where a user defines a
workflow. The first section, called ``Options'', specifies high-level options
that are used by the \texttt{hera\_opm} package in the construction of the
executable script files. For instance, it allows the user to provide a default
number of processors or memory quantity for batch jobs, or an Anaconda
environment that should be activated before the execution of the work
script. These options are fully enumerated in the documentation of the
\texttt{hera\_opm} package. Following the Options section, the ``WorkFlow''
section provides the main execution order for the pipeline. Further below,
options are specified for each step, including the arguments that are passed in
to the execution script. A corresponding batch job is generated for each step
for each specified input file, with appropriate substitutions made for the file
name and any necessary command line arguments. These batch job files are
relatively simple, and only consist of: (i) creating an appropriate environment,
(ii) calling the shell script which performs the required task, (iii) generating
a ``success'' file if execution succeeded or a ``failure'' file if not. The
\texttt{makeflow} program uses the existence of the ``success'' files to signify
the task is completed and execution should continue on to the next job. If not,
the task will be retried a certain number of times before being abandoned.

The shell scripts associated with each task follow a particular convention of
being named for their corresponding step. For example, the script for running
the \texttt{SETUP} task is called \texttt{do\_SETUP.sh}. These scripts in turn
can call scripts that perform the actual work associated with the task, such as
Python scripts, C programs, or command-line utilities. The path to these scripts
is specified in the configuration file, and the existence of these scripts is
treated as a pre-requisite for the step, as a means of ensuring they exist
before the pipeline is started. In the specific use case of invoking Python
scripts and libraries, the \texttt{hera\_opm} infrastructure allows for
activating a target Anaconda environment, which can be specified as part of the
configuration file. Besides this feature, the software environment is managed by
the user and inherited from the execution shell.

The input for generating a pipeline are a configuration file and a list of data
files to be processed. Based on the contents of the configuration file and the
data files, the \texttt{hera\_opm} package will generate a workflow script that
can be parsed by \texttt{makeflow} and a series of shell scripts that will be
run. In general, there is one shell script per step, per file, though there is
support for processing ``chunks'' of input files as part of a single job. Once
the workflow and shell scripts are generated, \texttt{makeflow} is used to
manage the execution of the shell scripts. Successful completion of a shell
script generates a placeholder output file that signifies successful completion,
at which point \texttt{makeflow} launches the next job in the workflow. Once all
the jobs have been completed successfully (or alternatively, if a job has failed
more than an allowed number of times), \texttt{makeflow} halts execution of the
pipeline as ``finished.''

Although there is no formal requirement on the amount of work allowed as an
individual step in the pipeline, in general tasks are encouraged to be as atomic
as possible. For example, one might want to generate RFI flags for each
\textit{observation}, harmonize them into a single ``per-observation'' mask, and
then apply them to all raw data products. Rather than performing all such tasks
as a single job, the throughput of the total workflow may be improved by
dividing the work into three separate jobs. Such an approach helps take
advantage of the parallelism that execution on multiple compute nodes (or
multiple cores on a single node) can offer. It also helps identify portions of
the pipeline prone to failure more easily, and potentially resume stalled jobs
after fixing a bug in a specific step. The atomic approach can have drawbacks,
though if there is significant re-use of data between steps. For steps that are
dominated by I/O rather than compute capability, it may actually be beneficial
to perform several closely related steps as a single operation in the pipeline
rather than using an atomic approach. Another concern is that including too many
atomic steps can generate a workflow graph that becomes too large to compute in
practice, so a trade-off may be necessary for a large number of workflow steps
and input files. In general, profiling of various pipeline steps is
beneficial. We discuss some performance profiling tools used as part of RTP
below in Sec.~\ref{sec:monitoring}.

One feature worth mentioning in the description of the workflow are the
``setup'' and ``teardown'' steps. These are operations that are meant to be
performed once at the beginning and end of the workflow run, respectively. For
instance, the setup step may pull and install the latest versions of reference
data for software packages, and the teardown may remove all orphaned files not
cleaned up as part of the workflow. Their presence in the configuration file is
optional, though if they are present, the setup and teardown steps must be the
first and the last ones, respectively. As with other steps in the workflow, it
is possible to specify arguments for them.

\subsection{On-Site Analysis Workflows}
\label{sec:onsite}

\begin{figure}
  \centering
  \includegraphics[width=0.48\textwidth]{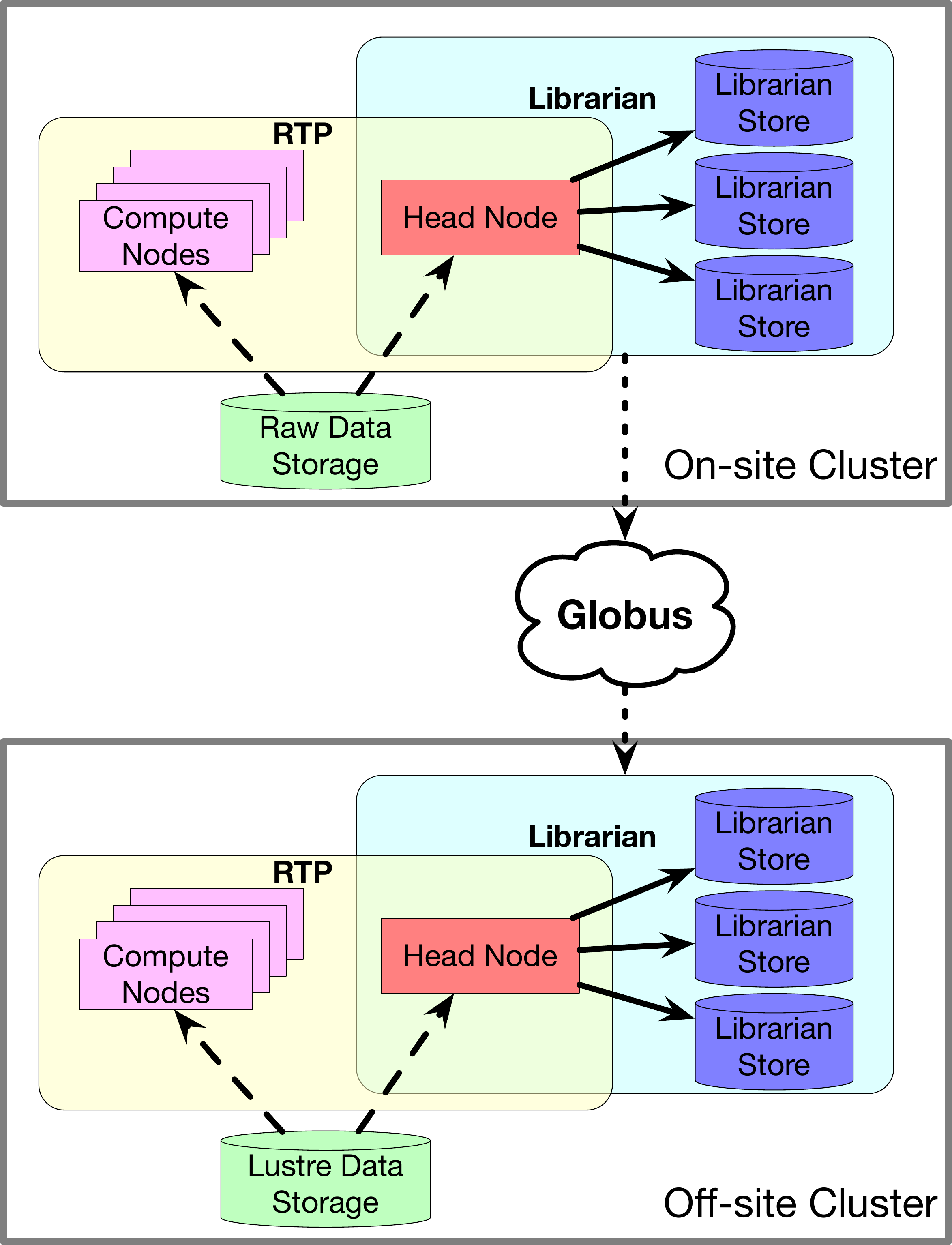}
  \caption{A schematic diagram showing the various components of computing
    hardware in the on-site HERA cluster, and how the Librarian and RTP systems
    are distributed across them. In addition to the on-site cluster, an off-site
    processing facility is shown, which represents the cluster at the National
    Radio Astronomy Observatory (NRAO). Off-site processing at the Ilifu
    research cloud infrastructure in South Africa is handled
    similarly. Transfers between Librarian instances is facilitated by Globus
    (\url{https://www.globus.org/}), a service that facilitates large-scale data
    transfer between computing facilities. See Sec.~\ref{sec:offsite} for
    further discussion of the interconnection of multiple sites.}
  \label{fig:qmaster}
\end{figure}

The on-site data analysis for HERA has the constraint of looking at data in
``real time'', i.e., shortly after they are taken. In addition to this time
constraint, the data volume must also be reduced by a factor of 100 or more in
order to move it within the network bandwidth available. We do not explore in
detail the analysis steps taken, but instead focus on some of the requirements
of various steps to highlight several features of the \texttt{hera\_opm}
package. The generation of a pipeline workflow file and shell scripts is
performed automatically by a monitoring system, after which the execution of the
pipeline begins.

As mentioned above, the \texttt{makeflow} program is used to control the overall
flow of the pipeline steps. One important feature of \texttt{makeflow} is its
ability to interface with several different cluster resource managers. For the
on-site pipeline in HERA, we make use of Slurm, which helps distribute the work
of various tasks across a heterogeneous series of compute nodes. In particular,
several analysis steps can take advantage of graphics processing units (GPUs) to
accelerate calculations, and assigning these nodes to separate partitions in the
Slurm cluster allows for specific tasks to be assigned to them. These computing
options are specified in the configuration script (as for the \texttt{XRFI} step
seen in Listing~\ref{list:config}), and are handled by Slurm.

Due to the processing constraints imposed by the data rate, the correlator
writes files that are only 16 seconds in length. For some steps, such as the RFI
flagging done in the \texttt{XRFI} step, analyzing many contiguous time samples
at once is scientifically important. In the case of RFI, multiple time samples
help define a ``baseline'' for the data, so that flagging anomalous
time-variable behavior is statistically simpler. For situations such as these,
the RTP system allows the user to specify options for handling an automatic
partitioning of the data into time-contiguous chunks. These options include: the
number of files to include in a single time-contiguous chunk (up to and
including all files in the workflow); the stride between these chunks (which in
general can be different than the number of files in a chunk, if for example one
wants to run analysis on overlapping chunks of data); whether the chunks are
symmetric about a central file when generating groups, or are simply counted
off; and how to handle an incomplete final group of files that is not evenly
partitioned by the group size. The workflow designed by \texttt{hera\_opm}
accordingly tracks the files that belong to each group individually, so that
downstream pre-requisites are handled automatically and correctly.

Figure~\ref{fig:qmaster} shows a schematic diagram of how the on-site cluster is
organized, including the machines that run RTP and the Librarian system
(discussed further below in Sec.~\ref{sec:librarian}). The RTP system is
primarily concerned with running analysis, and makes use of the compute
nodes. These nodes are assigned to various partitions in the Slurm system, and
are available to run batch computation jobs. The head node serves as the
submission host for Slurm jobs, as well as generating the initial workflow job
using \texttt{hera\_opm} when new data have been generated. The main raw data
storage nodes are mounted on both the head node and each of the compute nodes
via NFS mount (shown as a dashed line in the diagram), which allows for sharing
data amongst the various nodes without the need to duplicate data. Intermediate
data products needed for further analysis are written to the raw storage
space. Output products earmarked for permanent storage and transfer back to US
clusters are uploaded to the Librarian. In general, tasks which upload files to
the Librarian are separate jobs in the workflow. As mentioned above in
Sec.~\ref{sec:rtp_architecture}, the steps in the workflow are made as atomic as
possible. For example, in the sample workflow shown in
Listing~\ref{list:config}, steps uploading data to the Librarian are included in
the workflow as separate tasks, as are analysis steps which feature different
access patterns through the data.

\subsection{Off-Site Post-Processing}
\label{sec:offsite}

An important feature of the \texttt{hera\_opm} module is its portability: in
principle, it is possible to run in any cluster environment that uses cluster
resource management systems, or even on personal laptops. Because the machinery
of \texttt{hera\_opm} relies only on the Python-based package and
\texttt{makeflow} for handling the computational flow, virtually any system
should be able to make use of the infrastructure. In particular, HERA has access
to a computing facility hosted by the National Radio Astronomy Observatory
(NRAO). This cluster has a very different set of computing resources available
compared to the on-site cluster, but the process of generating a workflow based
on inputs using \texttt{hera\_opm} is identical. The NRAO cluster uses the
TORQUE resource manager, which is supported as a \texttt{makeflow} batch
computing system. In processing the computing options in the configuration file,
\texttt{hera\_opm} is aware of differences in options between the two systems,
and so the user is not required to make any changes in the configuration file
apart from specifying a different queue system. When running the actual
workflow, these steps are executed in a similar fashion. Assuming the input
files and installed versions of software libraries are the same, then the output
will be identical. The support for running on multiple clusters easily allows
for portability of the entire analysis stack, and supports straightforward
verification of any analysis output products.

\begin{figure*}
  \centering
  \includegraphics[width=0.95\textwidth]{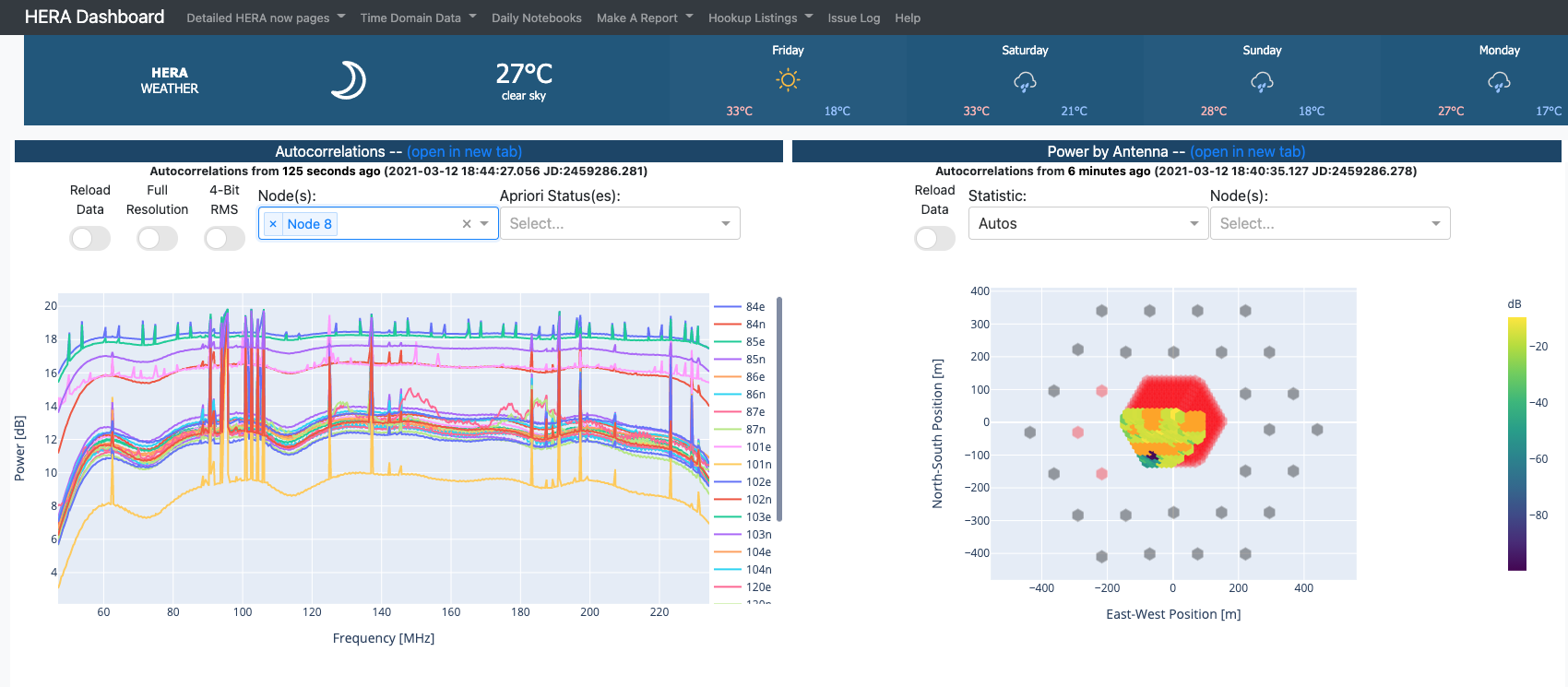}
  \caption{An example of the HERA dashboard. The top panel shows the current
    weather for the HERA site in the Karoo Desert of South Africa. The lower
    left panel shows live autocorrelations for the telescope, and allows the
    user to select specific antennas or sections of the array to be
    displayed. The lower right panel shows the status of individual antennas as
    part of the HERA array. In addition to the panels shown here, there are
    several other pages that show the status of other components of the
    telescope. The website is accessible both on-site and off-site which allows
    remote observers to understand the current state of the system.}
  \label{fig:hera_dashboard}
\end{figure*}

The portability of the \texttt{hera\_opm} package also allows for breaking up
portions of the RTP system across different machines. Although currently all of
the processing is run as part of the on-site cluster, the ability to easily and
reliably make use of multiple computing environments opens the possibility of
using multiple clusters for different portions of the RTP tasks. For instance,
there are significant computing and processing resources available as part of
the Ilifu research cloud infrastructure\footnote{\url{http://www.ilifu.ac.za/}}
of South Africa, which supports MeerKAT and SKA Pathfinder telescopes in their
processing needs. Although additional infrastructure is required to ensure
coordinated execution on various remote hosts, distributing the overall resource
requirements across clusters is an overall beneficial approach to ensuring the
required computing is accomplished in the requisite time window.

\subsection{Pipeline Monitoring}
\label{sec:monitoring}
As discussed in Sec.~\ref{sec:rtp_architecture}, RTP as a system is primarily
concerned with executing a series of jobs in a user-defined workflow. However,
there are several monitoring tools built into RTP that help remote observers
monitor the overall state and health of the system. In this section, we briefly
cover some of the tools developed to make this monitoring possible. We make use
of PostgreSQL\footnote{\url{https://www.postgresql.org/}} (PSQL),
InfluxDB,\footnote{\url{https://www.influxdata.com/}}
Chronograf,\footnote{\url{https://www.influxdata.com/time-series-platform/chronograf/}}
and Grafana.\footnote{\url{https://grafana.com/}} These subsystems allow for
externally monitoring pipeline status without the need to directly log into the
on-site system, and provides users with time-series data regarding key metrics
related to the system. With the use of Chronograf and Grafana, metrics and
statistics are plotted in real time without need for direct on-site access.

The key metrics, such as information about the state of processing systems, are
stored in PSQL and InfluxDB servers running on-site. The database is populated
by a system agent running on each computation host. This agent provides
information such as the computational load, memory usage, and disk capacity as a
function of time which are stored in the InfluxDB. It also contains key metrics
related to the state of the correlator which are sent to the PSQL database. This
information on the health of the processing system is also tracked by the
Monitoring \& Control subsystem, though a full description of this system is
outside the scope of the current discussion. As it pertains to the status of the
RTP system, these metrics are important for making sure the system is operating
as intended, and has not stalled. The full data rate of 50 TB per night
necessitates decreasing the total volume of data that are stored indefinitely,
and so processing must continue with high reliability. Having tools for
assessing and measuring these processes is essential to ensuring that HERA can
deliver on its science goals.

Figure~\ref{fig:hera_dashboard} shows a sample of the HERA dashboard, which
contains information about the status of the telescope itself and data
processing. On the left, this particular page shows the autocorrelation spectra
from active antennas in near-real time, which can be useful for identifying
poorly performing components. The user can easily select particular antennas to
show, or get information for specific sections of the array. The right part of
the page shows the layout of the entire array, with individual antennas
color-coded based on their performance. In this particular image, the large
number of red antennas denote ones which have not yet been connected, as HERA is
still partially under construction at time of writing. In addition to these
high-level views of the system, additional pages display more detailed
information about logs from specific machines, the historical status of
individual antennas, the fraction of files that have been successfully processed
by RTP and moved to the USA for further analysis, and more. Taken together, this
monitoring system allows all members of the collaboration an easy and
straightforward way to ensure the array is behaving as expected, which increases
the amount of science-quality data generated by the telescope.

\section{Long Term Storage and Data Transfer: The Librarian}
\label{sec:librarian}
We have also developed a new system which handles large data volumes and
automated data processing specifically addressing several needs not covered by
existing systems, as described in Sec.~\ref{sec:system_design}.  The Librarian
system addresses these requirements with a database backed program that records
file locations as part of metadata, handles moves, and provides an API suitable
for queries by automated pipelines or manual user control via a web page. As
with the RTP system, it is written in Python, which allows for easily installing
on a wide variety of systems. Although execution of commands on remote hosts
assumes a Linux-based environment, these portions of the code could be easily
replaced with analogous commands for other systems.

The HERA RTP system interacts closely with the Librarian, which provides
long-term storage and data management capabilities for the HERA
collaboration. At scale, HERA is projected to write 50 TB of data spread across
50,000 unique files each night. The Librarian system is designed to keep pace
with the required data ingest operations, provide an interface for access to the
accumulated HERA data set, and implement data transfer allowing members of the
international HERA collaboration to perform a wide range of analyses at their
local institutions.

\subsection{Architecture}
\label{sec:lib_arch}

The Librarian system is composed of a loose federation of
independently-operating \textit{sites}. Each site hosts a freestanding Librarian
server, which manages a database of file information, and \textit{storage nodes}
where file contents are actually stored. Each server exposes a JSON-based HTTP
API that can be accessed by \textit{clients} using several means: a Python
library (the \texttt{hera\_librarian} package), a command-line client, or an
HTML web interface. Clients may be humans, RTP tasks and other local automated
systems, or remote services such as other Librarian servers performing
site-to-site synchronizations.

A key element of the Librarian design is its multi-site architecture. In the
HERA collaboration, the two main Librarian sites are the observatory itself,
where raw data are generated, and NRAO Domenici Science Operations Center (DSOC)
in Socorro, NM, USA, which provides computing support to the collaboration. Each
site runs its own Librarian server and mostly operates independently. At the
same time, different sites can communicate and send data to one another as
needed. In routine operations, transfers from South Africa to the USA happen on
a daily basis as new data are recorded. However, if anything prevents
site-to-site communication, such as loss of internet connectivity at the Karoo,
users experience no issues besides a lack of the newest data. In terms of the
CAP theorem \citep{gilbert2002}, which refers to the choice between
\textit{consistency} or \textit{availability} in the presence of
\textit{partition tolerance} (i.e., network transmission failures), the
Librarian prioritizes availability far ahead of consistency.

In typical cases, data flow through the HERA Librarian system in a radial
pattern: raw data are generated in South Africa and transferred ``outward'' to
the USA and other processing sites. Despite this, the Librarian system is
designed to support more varied flow topologies. For instance, an analysis of
raw data might be run in the USA to generate new calibration files, which are
then synchronized to South Africa to be used in RTP processing, creating a
circular data flow.

The existence of multiple copies throughout the global system of Librarian
instances poses a variant of the problem solved by version trackers like
\texttt{git} or \texttt{mercurial} which we address with a similar solution to
the one used by \texttt{git}: a two-tiered, append-only data model. This model
has three main components:
\begin{enumerate}
\item The Librarian distinguishes between \textit{files} and \textit{file
    instances}. In the Librarian, a ``file'' is defined by its metadata: name,
  size, MD5 digest, creation time, and so on. A ``file instance'' is an actual
  copy of the file on a storage node. A Librarian server may hold zero, one, or
  many instances of a file.
\item All file metadata and file instance data are immutable.
\item File names are globally unique. Files can therefore be uniquely identified
  by their names.
\end{enumerate}
Each Librarian server's database of files may therefore adopt append-only
semantics. If files are to be sent from one Librarian site to another, the
metadata are easy and inexpensive to synchronize. Actual file instances are more
expensive to synchronize, because they are generally much larger, but their
immutability means that such synchronization can always eventually succeed.

In this model, the convention for choosing file names is essential to the
system's stability. While a single Librarian server can refuse to ``mint'' a new
file record with a name that it knows to have already been taken, if two
different servers mint files with the same name, there is presently no mechanism
to reconcile their disagreement. In the experience of the HERA collaboration,
file name clashes have never been a problem because the HERA Librarian sites
generate different categories of data files (raw data, calibrations, data
releases) and each such category has a distinctive file naming scheme. This
convention maps well to real-world operations of scientific collaborations.

A ``file'' in the Librarian does not need to be a single traditional file. The
Librarian software can handle directory trees as single coherent ``files'' for
its indexing purposes, adopting a simple prescription for generating a
reproducible MD5 digest. Though originally born out of necessity because several
radio astronomy data formats are directory-based, this behaviour is useful for
combining multiple files into a single unit before uploading to the Librarian,
without the need to join the files together using \texttt{tar} or a similar
command line utility.

Finally, each Librarian server also maintains a loosely-structured log of
\textit{file events}. These record simple operational data (e.g., ``a new
instance of this file was created on storage node X'') and housekeeping
information such as records of successful ``standing order'' transfers as
described below. File events are not synchronized between sites.

The Librarian server uses the
Tornado\footnote{\url{https://github.com/tornadoweb/tornado}} framework for HTTP
services and keeps all information in a PSQL relational database. It is
statically configured with knowledge of local storage nodes, which are typically
separate machines equipped with RAID arrays or other large-format
devices. Storage nodes must be accessible over \texttt{ssh}, and must have some
small Librarian utility programs installed, but do not need to run any special
server software. Data transfers in and out of the Librarian system occur
directly between clients and storage nodes: the server determines the commands
necessary to move files between stores and executes them via remote execution on
stores, rather than launching the transfers directly on the server.

\subsection{Data Ingest}
\label{sec:data_ingest}

\begin{figure*}[t]
  \centering
  \includegraphics[width=0.95\textwidth]{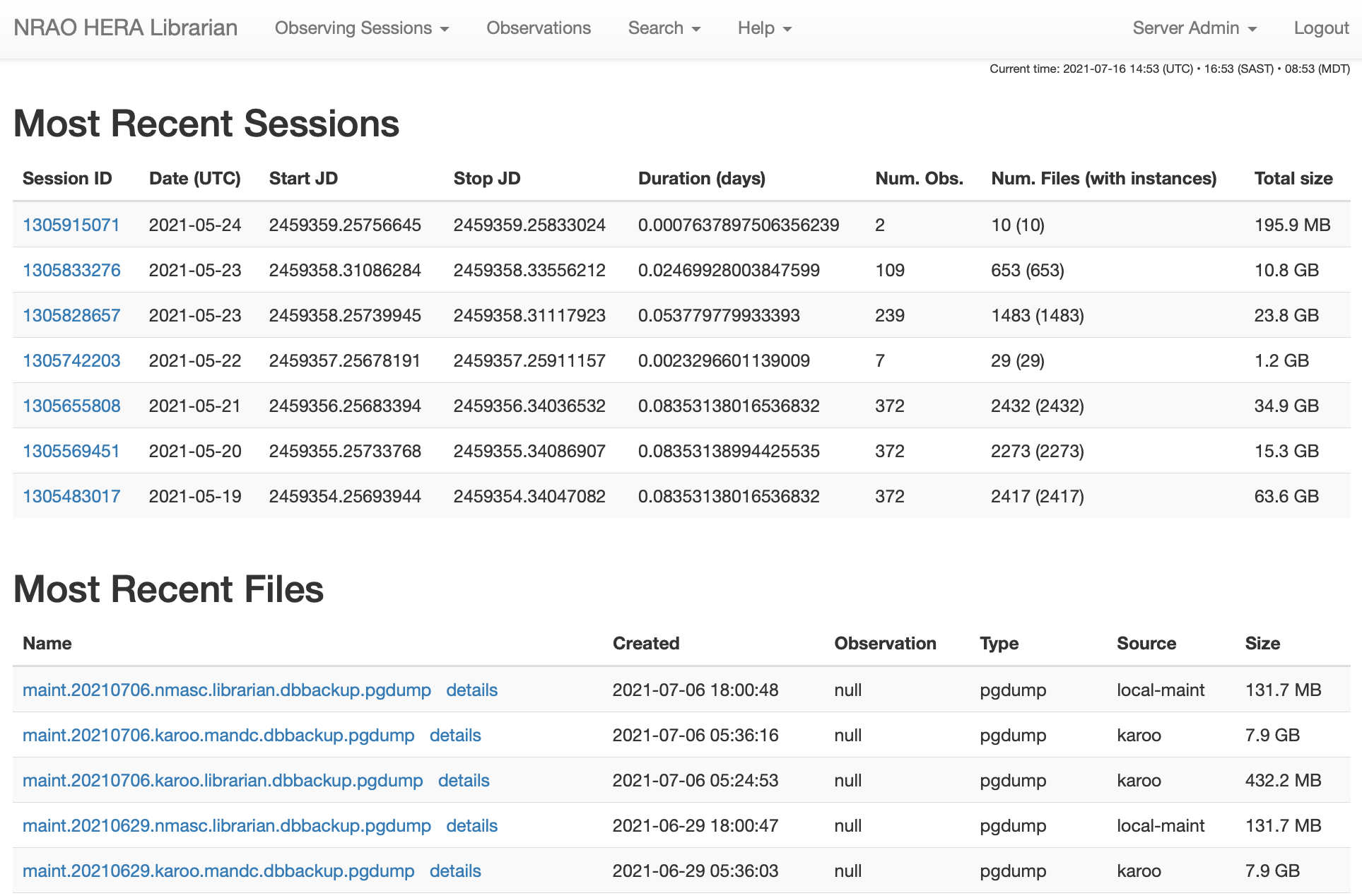}
  \caption{A sample of the Librarian web interface. A list of recent observation
    sessions is at the top, with more detailed information below. Clicking on
    the number identifying an observation session takes the user to a page that
    provides additional details about the data, with similar behavior for the
    identifying numbers of individual files. From here users can download
    individual files or (more commonly) make a copy of datasets on fast cluster
    computing storage from the slower main Librarian storage.}
  \label{fig:lib_login}
\end{figure*}

Data ingest into the Librarian is performed in a two-step process to ensure
reliable operation. In the first phase, the client computes key metadata such as
file size and MD5 digest and reports them to the server. If the upload is
allowed, the server creates space in a ``staging area'' on one of the storage
nodes and instructs the client to upload its data. In the second phase, the
client uploads the data and, if the upload completes successfully, notifies the
server of that fact. The server (via support tools installed on the storage
node) verifies that the uploaded data agree with the client's claims and, if so,
atomically adds them to its data holdings. Because the Librarian has made its
own copy of the file, the user is free to choose whether to delete the original
copy upon the successful completion of an upload operation or retain a local
copy.

While some file metadata are fully generic (e.g., size), the Librarian tracks
additional data that are more specific to its use as a astronomical data
manager. Of particular note is that each file is associated with an
\textit{observation} record defined by a start time (expressed as a Julian date,
JD) and a duration. By convention, the JD is saved in the filename of raw HERA
data, so derived data products which share a common prefix with these raw data
can be matched to the same observation. When a derived data product is created
from inputs corresponding to more than one \textit{observation}, it is generally
named according to the earliest \textit{observation} of all of its inputs. While
simple, this grouping scheme has worked well for the HERA collaboration's usage
patterns.

The Librarian has special support for one further layer of grouping. The user
may instruct the Librarian server to group observations into \textit{observing
  sessions}, each session ideally corresponding to a contiguous block of
\textit{observations} spanning a full night of observing. This assignment is
done automatically and progressively: when so instructed, the server determines
sessions from the start times and durations of all \textit{observations} that
have not already been assigned to a session. Typically, this grouping is
performed automatically at the end of uploading a night's worth of observations
to the Librarian. These observing session groupings are then used by the RTP
system for defining a workflow and performing analysis.  Observation and
observing session parameters are immutable once assigned.

\subsection{Searches}
\label{sec:lib_searches}

The Librarian is equipped with a simple search framework. Besides allowing human
users to explore the Librarian data holdings, the search infrastructure is also
intended to be a useful building block for system automation tasks, such as the
``standing order'' system which moves data between sites. Standing orders are
defined in terms of search queries and evaluated periodically (typically a few
times per hour). Files that match the search criteria for standing orders are
added to a list of transfers to be performed. This mechanism allows for
automatic data transfer between, e.g., the on-site Librarian and the one at
NRAO, as shown diagrammatically in Figure~\ref{fig:qmaster}.

Searches in the Librarian are expressed as JSON documents that are transcribed
into SQL queries. Searches are represented in the JSON document as trees of
clauses composed using standard boolean operations. Most search clauses map
directly into SQL \texttt{SELECT} terms: for instance, a \texttt{"name-like"}
JSON term is translated by the Librarian's search API into a \texttt{name LIKE}
query in the resulting SQL. Searches may query the Librarian's tables of files,
observations, or observing sessions. Search results for file queries can yield
structured file metadata, lists of file names, or lists of file instances. For
HERA, these metadata include the time and duration of individual observations,
which ``observing session'' an individual observation belongs to, and the local
sidereal time of an observation.

Certain search clauses can be fairly expensive to evaluate. For instance, one
can search on the number of files associated with a given observing session,
which is useful for automating checks for serious failures in the data recording
system. The Librarian computes this number on-the-fly every time that it is
needed, performing a query on its backing database that must join over the
tables of files, observations, and observing sessions.

\subsection{Site-To-Site Data Transfers}
\label{sec:data_xfer}

Data transfers between Librarian sites follow the same two-phase process as is
used for intra-site, client-to-server, transfers. Each Librarian server may be
configured to have ``standing orders'' to automate data transfers. Each standing
order is defined by a file search query and a destination site. The server
periodically evaluates the query and identifies any files that have not yet been
sent to the destination. Transfers of these files are then initiated. These
uploads are performed with modest parallelization because multiple streams can
make use of the available network bandwidth more effectively than a single
one. When each file has been successfully transferred, a record of the transfer
is made in its event log, preventing future upload attempts. As with all aspects
of the Librarian, this design of this feature is intended to provide reliable
operation in the face of unreliable network availability between sites.

Inter-Librarian data transfers can make use of Globus, a service that
facilitates large-scale data transfer between computing facilities. Globus
primarily serves facilities that support scientific research, though it also
supports transfer to commercial computing infrastructure like Amazon Web
Services (AWS\footnote{\url{https://aws.amazon.com}}). Many supercomputing
facilities are already established as Globus endpoints, which allows users to
tap into their infrastructure. In addition to tools that automatically adapt for
various network configurations, Globus also provides detailed information on
file transfer status, robust logs for records, and notifications on successful
(or unsuccessful) transfers. Setup of a Personal Globus Endpoint on the
Librarian server is straightforward, and only requires a few additional entries
in the Librarian configuration file. In the case of HERA, the NRAO computing
facility is a Globus endpoint, and the computing facility on-site makes use of
Personal Globus Endpoints to initiate transfers to the NRAO facility. In this
way, HERA data transfers are able to offload some of the logistical issues
related to data transfer by using Globus.

\subsection{Client Interfaces for Access to the Librarian}
\label{sec:lib_interface}

Each Librarian server supports client access through a command-line tool, a
Python library, and an HTML web interface. Most client access is ultimately
provided through the \texttt{hera\_librarian} Python package, which has few
dependencies and so can be easily installed by the end user. The user configures
the client with knowledge of one or more Librarian servers through a small
configuration file. The \texttt{hera\_librarian} module then provides fairly
direct access to the JSON APIs provided by the remote servers. For example, a
script might implement a slightly more sophisticated version of the ``standing
order'' system, query a local Librarian server for a particular series files, or
apply a filter that is too complex to implement server-side. More concretely,
the HERA operations team has used the Python API to query a remote Librarian to
check which of the remaining files it holds, and then initiating a transfer from
the local Librarian server to the remote one. The Python API provides the most
powerful and flexible way of interacting with a Librarian server, which makes it
a valuable tool when using the Librarian in practice.

The \texttt{hera\_librarian} Python package provides a command-line tool,
\texttt{librarian}, that provides a more convenient interface to the most common
Librarian access tasks, such as uploading new files to the librarian and
searching for existing files. For this latter task of searching for files, only
a few key bits of information are provided, such as the file name and size on
disk. For more detailed information, the user must use the Python API or the
web-based interface. Users can install the \texttt{hera\_librarian} package in a
local Python environment, and connect to either local or remote Librarian
servers. The information details of accessing servers is handled via a per-user
configuration file, which may require port forwarding via an SSH connection to
access servers behind a remote firewall.

Finally, each Librarian server provides an HTML user interface along with its
JSON API. While a rudimentary authentication scheme is implemented, the primary
means of access control is SSH authentication. HTTP interfaces to the Librarian
are not exposed on the open Internet: as with using the Python API, SSH
port-forwarding may be required to access the web server.

Figure~\ref{fig:lib_login} shows a sample of the Librarian web interface. At the
top is a summary of the most recent Observing Sessions, including their Julian
Date and the number of constituent Observations. Clicking on the Observing
Session identification number will take the user to a page with more detailed
information. Below the section of Observing Sessions, the Librarian displays the
most recent files added to it. This list merely provides a quick overview of the
most recent files, and is not intended to be an exhaustive list of all files in
the Librarian. The web interface also has a search page, where the user may
construct a query matching specific files more easily. From here users can
download individual files or (more commonly) make a copy of datasets on fast
cluster computing storage from the slower main Librarian storage.

The Librarian web interface is intentionally basic, because the expectation and
intention is that day-to-day data analysis should be driven through command-line
access or Python scripts while on site the system is tracked by the
HERA monitor and control system. While the web interface is helpful for
administration and monitoring, with HERA's large data volumes and extremely
uniform data holdings, data retrieval should generally be done in some kind of
automated fashion, at the beginning of an analysis pipeline.

There is no particular requirement that Librarian client libraries must be
implemented in Python. The Librarian server APIs use standard HTTP and JSON
patterns, and so it would be straightforward to implement client software in
nearly any modern programming language.

\subsection{Docker Container}
\label{sec:lib_docker}

In addition to the ``bare-metal'' installation of Librarian described above, the
\texttt{librarian} package also supports installation as a series of
Docker\footnote{\url{https://www.docker.com/}} containers. Rather than using a
single monolithic container for the entire application, the different components
are separated into different images. This takes advantage of orchestration
software for container management to facilitate an overall higher uptime and
reliability. The Librarian package is composed of three containers: (i) the
backing PostgreSQL database, (ii) the Python-based Librarian application, and
(iii) a ``store'' server for saving data files. The latter two containers are
based off the same base image, which includes an installation of the
\texttt{hera\_librarian} package. The ``store'' image also runs an ssh server,
so that minimal changes are required to the traditional method of operation of
the Librarian. The networking interfaces between these containers is specified
in a \texttt{docker-compose.yml} file. A full series of instructions is
available in the repository describing the installation procedure. Note that the
SQL database and the location of the store server where data are saved should be
configured as Volumes, so that the data and metadata persist between Docker
image instances.

\section{Summary}
\label{sec:summary}

The HERA project has built on lessons from several projects and through a series
of iterations arrived at a system for processing and managing large data
volumes. Formative events in prior projects included loss of half a season of
data to accidental deletion, movement of data using frequent shipments,
inadvertent siloing of data inside an unscriptable archive system, and failure
of automated scripts leading to silent telescope shutdowns.  These problems
largely stemmed from a paucity of automation and are largely in the past. The
system described here is currently in operation at several sites around the
world processing 10 TB per day at peak operation---a rate which is expected to
grow to hundreds of TB per day in the coming months.

With the explosion of raw data that has come about in astronomy research, data
analysis and processing has become equally important. Developing flexible yet
reliable frameworks for analyzing and storing data are of the utmost importance
for providing trustworthy scientific results. We have presented the
RTP\cref{fn:rtp} and the Librarian\cref{fn:librarian} systems as they pertain to
data analysis in HERA, which has particularly strict requirements for processing
and data handling. However, these tools have been developed such that they are
modular and general enough to be applied in other astronomy contexts, such as in
experiments outside of radio astronomy or future projects like the SKA. The
source code underlying these systems is free and open source, and available on
GitHub. We hope that these tools may be useful to other collaborations or
projects as the volume of astronomy data will only increase in the future.

In the future, we plan to increase the amount of automation and transparency
that happen in the system. In particular, visualizing the status of the RTP
system through the monitoring website shown in Figure~\ref{fig:hera_dashboard}
will be essential for ensuring that data are successfully processed in a timely
fashion. We also plan to develop the multi-site processing paradigm discussed in
Sec.~\ref{sec:offsite} further, as the relatively limited computational
resources on site necessitate using additional compute resources. Throughout all
of these development efforts, the ability to test and deploy the code both
locally and at scale has been facilitated by continuous integration testing
principles (e.g., writing a robust series of tests which must pass before
changes can be committed to the main software repo), leading to higher quality
code and better reliability. Whether adapting these RTP and Librarian systems or
building bespoke ones, users facing similar design considerations are encouraged
to use robust testing frameworks for ensuring reliable operation of critical
software infrastructure components.

\section*{Acknowledgment}
This material is based upon work supported by the National Science Foundation
under Grant No. 1636646, the Gordon and Betty Moore Foundation through grant
GBMF5215 to the Massachusetts Institute of Technology, and institutional support
from the HERA collaboration partners. HERA is hosted by the South African Radio
Astronomy Observatory, which is a facility of the National Research Foundation,
an agency of the Department of Science and Innovation. Parts of this research
were supported by the Australian Research Council Centre of Excellence for All
Sky Astrophysics in 3 Dimensions (ASTRO 3D), through project number
CE170100013. G. Bernardi acknowledges funding from the INAF PRIN-SKA 2017
project 1.05.01.88.04 (FORECaST), support from the Ministero degli Affari Esteri
della Coorperazione Internazionale -- Direzione Generale per la Promozione del
Sistema Paese Progetto di Grande Rilevanza ZA18GR02 and the National Research
Foundation of South Africa (Grant Number 113121) as part of the ISARP
RADIOSKY2020 Joint Research Scheme, from the Royal Society and the Newton Fund
under grand NA150184 and from the National Research Foundation of South Africa
(grant No. 103424). P. Bull acknowledges funding for part of this research from
the European Research Council (ERC) under the European Union's Horizon 2020
research and innovation programme (Grant agreement No. 948764), and from STFC
Grant ST/T000341/1. J. S. Dillon gratefully acknowledges the support of the NSF
AAPF award \#1701536. N Kern acknowledges support from the MIT Pappalardo
fellowship. A. Liu acknowledges support from the New Frontiers in Research Fund
Exporation grant program, the Canadian Institute for Advanced Research (CIFAR)
Azrieli Global Scholars program, a Natural Sciences and Engineering Research
Council of Canada (NSERC) Discovery Grant and a Discovery Launch Supplement, the
Sloan Research Fellowship, and the William Dawson Scholarship at McGill.

\bibliography{mybib}

\end{document}